# Photoacoustic characterization of TiO$_2$ thin-films deposited on Silicon substrate using neural networks

**Djordjević KLj [1)\*], Markushev DK [2], Popovic MN[2], Nesic MV[1], Galović SP[1], Lukić DV[2], Markushev DD[2]**

(1) University of Belgrade, „VINČA" Institute of Nuclear Sciences - National Institute of the Republic of Serbia, University of Belgrade, PO box 522, 11000 Belgrade, Serbia
(2) University of Belgrade, Institute of Physics Belgrade, National Institute of the Republic of Serbia, Pregrevica 118, 11080 Belgrade (Zemun), Serbia

*katarina.djordjevic@vin.bg.ac.rs

**Abstract**

In this paper, the possibility of determining the thermal, elastic and geometric characteristics of a thin TiO2 film deposited on a silicon substrate, thickness 30 μm, in the frequency range of 20 to 20 kHz with neural networks was analyzed. For this purpose, the substrate parameters remained the known and constant in the two-layer model and nano layer thin-film parameters were changed: thickness, expansion and thermal diffusivity. Prediction of these three parameters was analyzed separately with three neural networks and all of these together by fourth neural network. It was shown that neural network, which analyzed all three parameters at the same time, achieved the highest accuracy, so the use of networks that provide predictions for only one parameter is less reliable.

Keywords: Thin-film, TiO2, photoacoustic, artificial neural networks, thermal diffusion, thermal expansion, inverse problem

## Introduction

In the last decade, TiO$_2$ has had a wide range of applications in coatings, medicines, plastics, food, inks, cosmetics, and textiles. In the form of thin-film TiO$_2$ has been used for a great variety of applications including photocatalytic degradation of organic pollutants in water as well as in air, dye-sensitized solar cells (DSSCs), anti-fogging, super hydrophilic,... [1-4].

Analysis of thin-films on the substrate has always been a challenge for photoacoustics (PA) because film thicknesses ranged from a few tens to several hundred nanometres. Depending on the thickness of the substrate (usually more than tens of microns), such film thicknesses are usually at the limit of experimental detection [5-8]. This means, for example, that the differences in the amplitude of the photoacoustic signals (PAS), generated by a two-layer sample (substrate + thin-film) in the case where only the thickness of the film is changed, are extremely small, which leads to a standard analysis of such a system to determine the effective values of the parameters two-layer system [9-17]. The analysis of such two-layer samples is theoretically demanding for determining the parameters of the thin-film, because no significant effect is expected from this thin layer in frequency measurement range, which is why the application of classical methods for inverse solving is unreliable. However, networks have been shown to work well in such situations, which is why we developed a network in an attempt to characterize a thin layer of $TiO_2$.

By establishing a good procedure of applying a neural network developed on a theoretical model [18-23] for accurate processing of experimentally recorded PAS silicon samples, an open photoacoustic cell (OPC) in the frequency domain [24-27] shows effective recognition and removal of instrumental influence [28-33], and detailed and precise characterization of the sample [34-39], leading to the best decision to select a well-photoacoustically characterized silicon sample as the substrate. The very procedure of deposition of a thin nano-layer of $TiO_2$ on Si is simple, and this structure is often used in MEMS and NEMS, which is why we will examine it photoacoustically.

In order to avoid the mentioned additional normalizations and the calculation of effective values, we resorted to the use of precise characterization of the silicon substrate sample, with the aim of using the theoretical two-layer model [40-56] for the precise determination of thin-film parameters. Neural networks were formed in the analysis of PAS generated from the Si substrate + $TiO_2$ thin-film system. We expect that the networks will easily recognize differences in signals caused by changing only film parameters (thickness, thermal diffusivity, coefficient of thermal expansion). We also expect that neural networks can easily determine the specified parameters of $TiO_2$ thin-film with satisfactory accuracy and reliability. To do this, we created a relatively small database of PAS for training and four types of networks; the first three, which serve for the individual prediction of only one parameter of the film, and the last type, which serve for the prediction of all three parameters simultaneously.

## Theoretical background

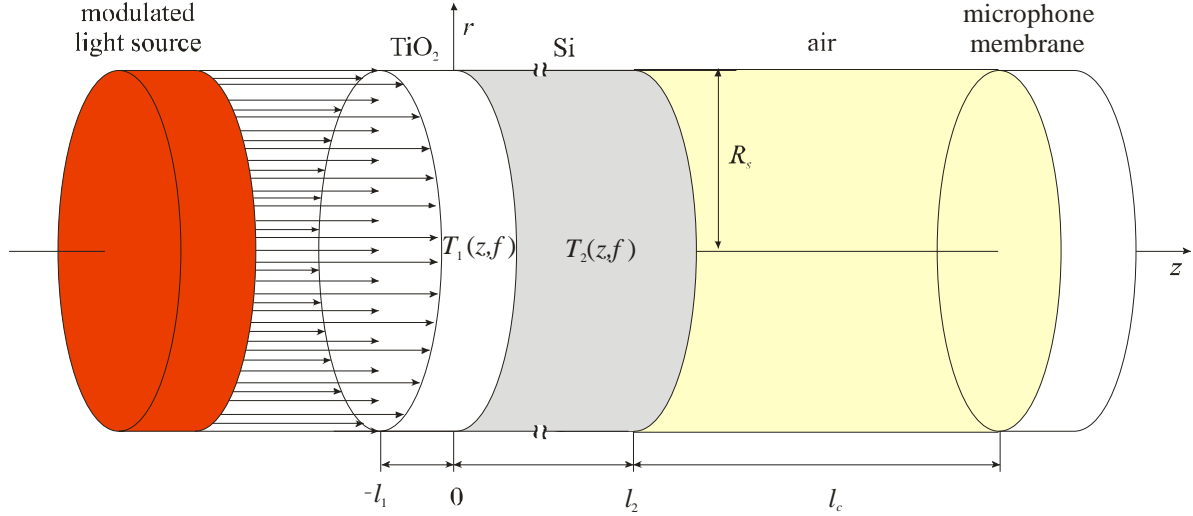

**Figure 1.** The simplest scheme of the two-layer sample irradiated by modulated light source. $l_1$ and $l_2$ ($l_1 \ll l_2$) are the thicknesses of the thin-film (TiO$_2$) and substrate (Si), respectively. $R_s$ is the sample radius, $T_1(z,f)$ is the temperature distribution in the thin-film and $T_2(z,f)$ is the temperature distribution in the substrate.

Illumination of the two-layer sample (Figure 1) with a modulated light source, wavelength 660 nm, produces a periodic change in the thermal state of both the film and the substrate. The TiO$_2$ layer is considered dielectric because there is no influence of photogenerated charge carriers due to larger energy gap of TiO$_2$ than photon energy of exciting beam while the exciting beam generates charge carriers in the semiconductor substrate. As a consequence of thermodynamic state changes, fluctuations exist in the air pressure behind the sample. Such fluctuations create different types of sound that result from thermal and elastic contributions from the sample (composite piston theory) that the microphone detects as a total PAS $\delta p_{\text{total}}(f)$, defined as [19-22,56-60]:

$$\delta p_{\text{total}}(f) = \delta p_{\text{TD}}(f) + \delta p_{\text{TE}}(f) + \delta p_{\text{PE}}(f), \tag{1}$$

where $f$ is the modulation frequency, and $\delta p_{TD}(f)$, $\delta p_{TE}(f)$, and $\delta p_{PE}(f)$ are the thermal diffusion (TD), thermo elastic (TE) and plasma elastic (PE) PA signal components, respectively. These components can be written as [19-22,56-60]:

$$\delta p_{TD}(f) = \frac{p_0 \gamma_g}{\sigma_g l_c} \frac{T_2(l_2,f)}{T_0}, \qquad (2)$$

$$\delta p_c(f) = \frac{\gamma p_0}{V_0} \int_0^{R_s} 2\pi r U_{z,c}(r,z) dr, \qquad c = \text{TE, PE} \qquad (3)$$

where $\gamma$ is the adiabatic constant, $p_0$ and $T_0$ are the equilibrium pressure and temperature of the air in microphone, $\sigma_g = (1+i)/\mu_g$, $\mu_g$ is the thermal diffusion length of the air; $l_c$ is the PA cell length, $T_2(l_2,f)$ is the dynamic temperature variation at the substrate rear (non-illuminated) surface [19-22,56-60] (see Appendix I), $V_0$ is the OPC volume, and $U_{z,c}(r,z)$ is the sample displacement along the $z$-axes (see Appendix II).

In PAs, the sound signals $\delta p_{total}(f)$ (Eq.(1)) are usually represented using its amplitudes $A(f)$ and phases $\varphi(f)$. Therefore, $\delta p_{total}(f)$ can be written as a complex number in the form:

$$\delta p_{total}(f) = A(f) e^{i\varphi(f)}, \qquad (4)$$

where $i$ is the imaginary unit. It is common to scale amplitude values for the application of neural networks in PAs in order to be comparable with the values of phase. A formula used for this purpose has a form:

$$A_{scale}(f) = 20 \log_{10} A(f). \qquad (5)$$

**Networks structure**

The structure of the networks used to characterize the thin-films on the substrate is shown in Figure 2. All networks used in this paper have the same structure: 2x72 input neurons (72 amplitudes and 72 phases) and 15 neurons in the hidden layer. The three networks, labeled NN1,

NN2, and NN3, have one neuron each in the output layer that serves to predict the $l_1$, $\alpha_{T1}$, and $D_{T1}$ thin-film parameters, respectively. The network designated as NN4 has three neurons in the output layer that simultaneously predict all three mentioned parameters. The bases formed for the training of the first three networks were made individually (Base 1, Base 2 and Base 3), while the training base NN4 (Base 4) was made by merging all three individual bases [61-64].

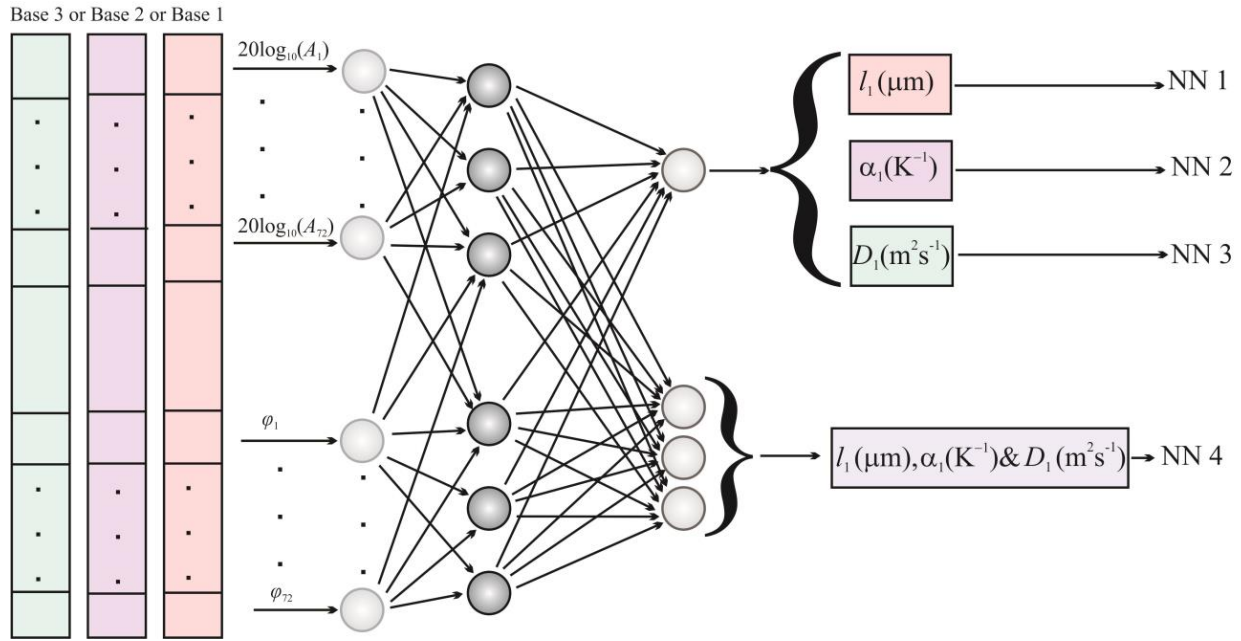

**Figure 2**. A representation of the structure of a single-layer neural network used for the training and prediction of $TiO_2$ thin-film parameters.

The training process neural network training on theoretical signals Bases 1-4, amplitude-phase characteristics, and the connection with the parameters of the thin-film is performed by an algorithm that uses statistical models of machine learning that enable prediction, Figure 2. In the prediction process, thin-film parameters are determined from the test signal or the experimentally recorded photoacoustic signal.

**Formation of the networks training bases**

The accuracy of the neural network largely depends on the selection of the basis for training, test and validation. The bases have been obtained numerically using Eq.(1-4). It is assumed that all of these signals are generated by the Si substrate + $TiO_2$ thin-film two-layer system presented in Figure 1. All formed bases consist of 41 PASs, one basic and rest of them obtained by changing in 10% the $TiO_2$ thin-film parameters. The basic signal parameters are presented in Table 1. Base 1 was formed for NN1 training, changing the thickness of $TiO_2$ film in the range of $l_2 = (475-525)$ nm with a step of 5 nm. Base 2 was formed for NN2 training, obtained by changing the coefficient of thermal expansion of $TiO_2$ film in the range of $\alpha_2 = (1.045-1.155) \cdot 10^{-5} K^{-1}$ with a step of $5 \cdot 10^{-8} K^{-1}$. Base 3 was formed for NN3 training, changing the thermal diffusivity of $TiO_2$ film in the range of $D_2 = (3.515-3.885) \cdot 10^{-6} m^2 s^{-1}$ with a step of $18.5 \cdot 10^{-8} m^2 s^{-1}$. Base 4 was formed for NN4 training, obtained by collecting 3x41 signals from all three previously mentioned bases. Since the signal changes in the bases are very small, we will show only Base 4 for clarity, bearing in mind that by one PAS we mean two curves presented to the networks: one for amplitude and another for phase (Eq.(4) and Figure 2).

By displaying the PASs of a silicon substrate of thickness $l_2 = 30 \mu m$ with different applied layers $l_1$ of $TiO_2$ thin-film, it is observed that there is no clear visual difference in the frequency dependence of the amplitudes, $A$, and that the factor of precise characterization by neural networks can be a visible difference in signal phases, $\varphi$, especially in the range from $10^3$ Hz to 20 kHz, Figure 3. The difference that exists in the phases is sufficient to train neural networks NN1-4, on the amplitude-phase characteristics and to correctly determine the parameters of a thin layer that is two orders of magnitude thinner than of his substrate.

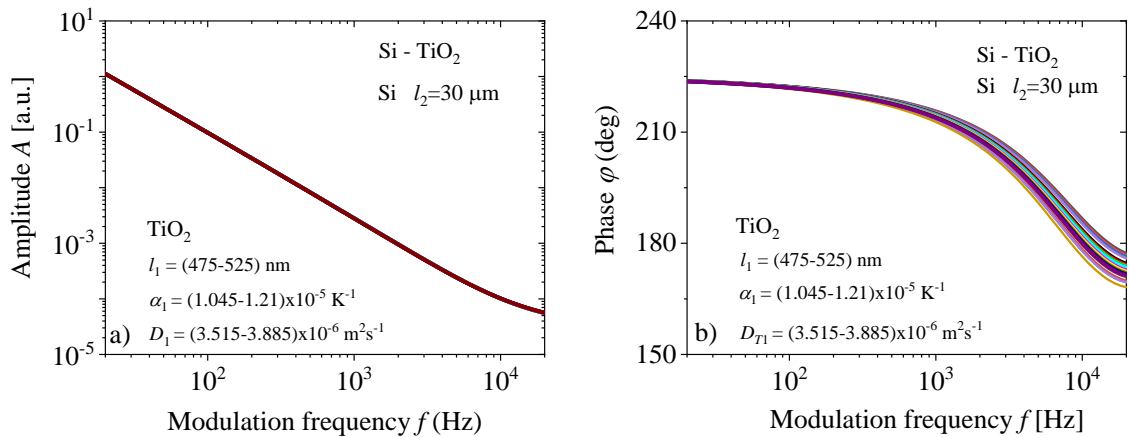

**Figure 3.** a) Amplitudes, $A$, and b) phases, $\varphi$, of the two-layer model: $TiO_2$ thin-films deposited on the Silicon substrate, obtained by changing parameters of the thin-film, diffusivity $D_1$, expansion $\alpha_1$ and thickness $l_1$.

**Table 1.** Values of basic parameter used for PA simulation $TiO_2$ thin-film deposed on Si substrate.

| parameters | labels | values |
|---|---|---|
| Air thermal diffusivity | $D_g[m^2s^{-1}]$ | $2.0566 \times 10^{-5}$ |
| Air thermal conductivity | $k_g[W(mK)^{-1}]$ | 0.0454 |
| Relaxation time of air | $\tau_g[s]$ | $2 \times 10^{-10}$ |
| Standard temperature | $T_0[K]$ | 300 |
| Standard pressure | $p_0[Pa]$ | $10^5$ |
| Air adiabatic index | $\gamma$ | 1.4223 |
| Si Thermal diffusivity | $D_{T2}[m^2s^{-1}]$ | $9 \times 10^{-5}$ |
| $TiO_2$ Thermal diffusivity | $D_{T1}[m^2s^{-1}]$ | $3.7 \times 10^{-6}$ |
| Si Thermal conductivity | $k_2[Wm^{-1}K^{-1}]$ | 150.0 |
| $TiO_2$ Thermal conductivity | $k_1[Wm^{-1}K^{-1}]$ | 11.0 |
| Si Thermal expansion coefficient | $\alpha_2[K^{-1}]$ | $2.6 \times 10^{-6}$ |
| $TiO_2$ Thermal expansion coefficient | $\alpha_1[K^{-1}]$ | $1.1 \times 10^{-5}$ |
| Si absorption coefficient | $\beta_2$ | $2.58 \times 10^5$ |
| $TiO_2$ absorption coefficient | $\beta_1$ | $1.8 \times 10^5$ |
| Si reflexing coefficient | $R_2$ | 0.3 |
| $TiO_2$ reflexing coefficient | $R_1$ | 0.2 |
| Si Young's modulus | $Ey_2$ | $1.37 \times 10^{11}$ |
| $TiO_2$ Young's modulus | $Ey_1$ | $1.0 \times 10^{11}$ |
| Si Poison coefficient | $v_2$ | 0.35 |
| $TiO_2$ Poison coefficient | $v_1$ | 0.30 |

## Results and discussion

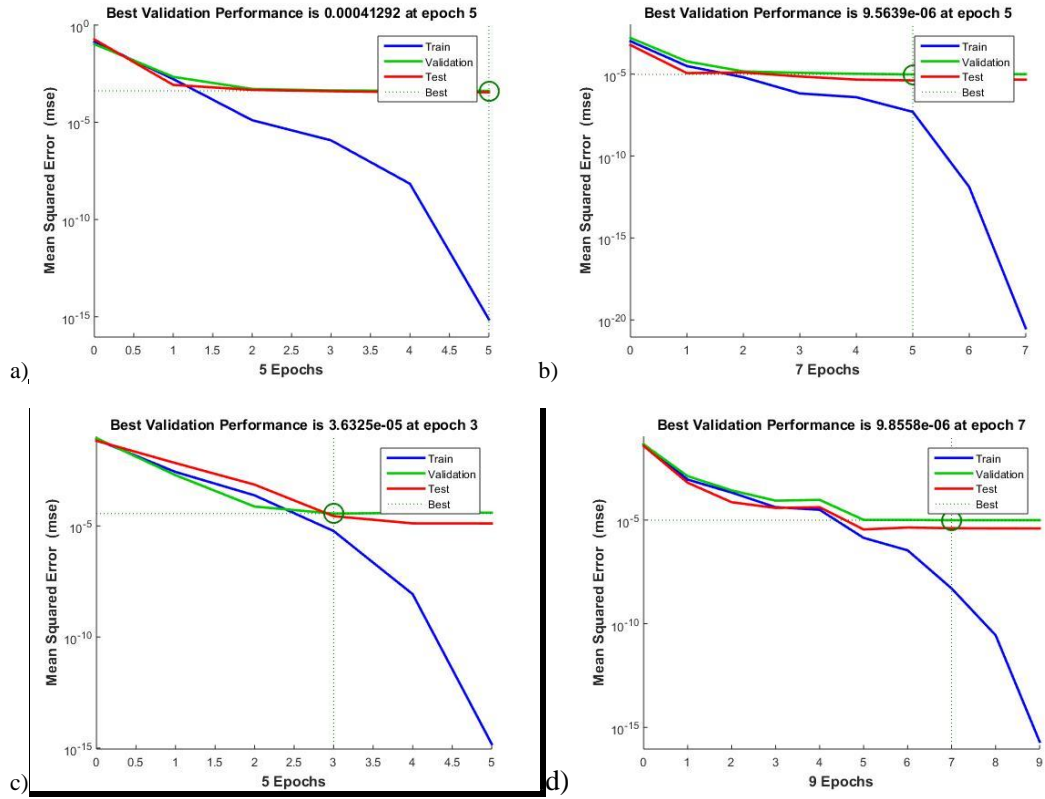

**Figure 4**. Network training: a) NN1, b) NN2, and c) NN3 for determining the parameters of thickness, expansion, and diffusivity of the $TiO_2$ thin-film, respectively, and d) NN4 for determining all three data, simultaneously.

The training results of the NN1-4 neural networks are given in Figure 4.a) -d), showing the Mean Square Error (MSE) of training, test and validation, depending on the number of epochs, and obtaining the best training performance. From each base for NN1-3 training, 4 signals were extracted for later testing. In the case of N4 training, 3x4 = 12 signals were also extracted from Base 4 for later testing. Network training interruption is activated by the deviation criterion of MSE training in relation to validation and test. The performance achieved by network NN1 is $4.1292 \times 10^{-4}$ in 5 epochs, network NN2 $9.5639 \times 10^{-6}$ in 5 epochs, and network NN3 $3.6325 \times 10^{-5}$ in 3 epochs and network NN4 $9.8558 \times 10^{-6}$ in 7 epochs. It can be seen by comparing these values that the best performance was obtained by NN4 and NN2 networks for determining all 3 parameters and expansion, respectively. The NN1 network obtained the weakest performance for determining the thin-film thickness parameter.

**Networks testing with in-step, out-step of PAS**

Table2. Relative (%) error prediction of TiO$_2$ thin-film parameters on 4 test PA signals that are in step by NN1, NN2 and NN3 networks.

| Type of network | NN1 | NN2 | NN3 |
|---|---|---|---|
| Base | 1 | 2 | 3 |
| parameter | $l_1^{NN1}$ | $\alpha_{T1}^{NN2}$ | $D_{T1}^{NN3}$ |
| TiO2 film no.1 | 0.4060 | 0.1041 | 0.3424 |
| TiO2 film no.2 | 0.1681 | 0.1270 | 0.1526 |
| TiO2 film no.3 | 0.1414 | 0.0690 | 0.2317 |
| TiO2 film no.4 | 0.0658 | 0.1583 | 0.0764 |
| Relative % error | 0.1953 | 0.1146 | 0.2008 |

Table3. Relative (%) error prediction of TiO$_2$ thin-film parameters NN4 on 4 signals from three bases "in-step" of training network.

| Type of network | NN4 | | | | | | | | |
|---|---|---|---|---|---|---|---|---|---|
| Base | 1 | | | 2 | | | 3 | | |
| parameters | $l_1^{NN4}$ | $\alpha_{T1}^{NN4}$ | $D_{T1}^{NN4}$ | $l_1^{NN4}$ | $\alpha_{T1}^{NN4}$ | $D_{T1}^{NN4}$ | $l_1^{NN4}$ | $\alpha_{T1}^{NN4}$ | $D_{T1}^{NN4}$ |
| TiO2 film no.1 | 0.7878 | 0.8782 | 0.4542 | 0.3579 | 0.1592 | 0.9610 | 0.2958 | 0.0951 | 0.0152 |
| TiO2 film no.2 | 0.0130 | 0.2941 | 0.3980 | 0.4126 | 0.4990 | 0.0564 | 0.0165 | 0.1059 | 0.2139 |
| TiO2 film no.3 | 0.0187 | 0.2002 | 0.1512 | 0.7414 | 1.1077 | 1.3738 | 0.0932 | 0.0016 | 03694 |
| TiO2 film no.4 | 0.1578 | 0.0588 | 0.2811 | 0.4206 | 0.8822 | 0.8822 | 0.1298 | 0.1278 | 0.0663 |
| Relative% error | 0.2443 | 0.3578 | 0.3211 | 0.4831 | 0.6434 | 0.8183 | 0.1325 | 0.0831 | 0.1661 |

As we said in the previous paragraph, four signals that did not participate in the training were separated from each training base of the NN1-3 networks. A similar thing was done with the training base for the NN4 network, from which 12 signals were separated and did not participate in the training. All four networks were tested with these "in-step" signals and the results of such a tests are shown in Table 2 and Table 3. Relative error predictions (%) presented in these tables, show that the most accurate networks are NN2 for prediction of $\alpha_{T1}$ and NN4 for prediction of $D_{T1}$.

Our next step is to check the quality of the prediction of neural networks with "out-step" signals - signals outside the training step but within the framework of parameter changes. For this purpose, 12 signals were randomly generated, four for each changed parameter $l$, $\alpha_T$ and $D_T$ individually. The prediction results for all four networks are given in Table 4 (N1-3) and Table 5 (N4). It is interesting to note that the NN1 network gives the worst prediction of sample thickness, while the NN4 network gives relatively satisfactory predictions for all three parameters.

**Table 4.** Relative (%) error prediction of TiO$_2$ thin-film parameters NN1-3 on 4 signals from three bases "out of step" of training network.

| Type of network | NN1 | NN2 | NN3 |
|---|---|---|---|
| parameter | $l_1^{NN1}$ | $\alpha_{T1}^{NN2}$ | $D_{T1}^{NN3}$ |
| TiO2 film no.1 | 2.4890 | 0.0186 | 0.0777 |
| TiO2 film no.2 | 2.4584 | 0.0293 | 0.0927 |
| TiO2 film no.3 | 5.4138 | 0.0011 | 0.2593 |
| TiO2 film no.4 | 4.8427 | 0.0031 | 0.0116 |
| Relative % error | 3.8099 | 0.0130 | 0.1103 |

**Table 5.** Relative (%) prediction error of TiO$_2$ thin-film parameters by NN4 for 4 signals "out of step".

| Type of network | NN4 | | | | | | | | |
|---|---|---|---|---|---|---|---|---|---|
| Base | 1 | | | 2 | | | 3 | | |
| parameter | $l_1^{NN4}$ | $\alpha_{T1}^{NN4}$ | $D_{T1}^{NN4}$ | $l_1^{NN4}$ | $\alpha_{T1}^{NN4}$ | $D_{T1}^{NN4}$ | $l_1^{NN4}$ | $\alpha_{T1}^{NN4}$ | $D_{T1}^{NN4}$ |
| TiO2 film no.1 | 0.1184 | 0.0422 | 1.33552 | 0.0173 | 0.0081 | 0.0164 | 0.0070 | 0.0049 | 0.0245 |
| TiO2 film no.2 | 0.0422 | 0.0116 | 1.3516 | 0.0055 | 0.0183 | 0.0153 | 0.0182 | 0.0140 | 0.0104 |
| TiO2 film no.3 | 0.0066 | 0.0599 | 1.3880 | 0.0080 | 0.0058 | 0.0270 | 0.0097 | 0.0215 | 0.0104 |
| TiO2 film no.4 | 0.1213 | 0.0044 | 1.3685 | 0.0781 | 0.0225 | 0.0520 | 0.0245 | 0.02385 | 0.0059 |
| Relative% error | 0.0721 | 0.0368 | 1.3658 | 0.0272 | 0.0137 | 0.0277 | 0.0149 | 0.0161 | 0.0128 |

**Networks testing with experimental signals**

The final part of our analysis is to test the ability to predict our networks on experimental signals. For this purpose, we measured, by the standard method of an OPC, the frequency response of a circular plate of a two-layer sample (silicon + TiO$_2$). Amplitudes and phases of the measured response (red stars) are shown in Figure 5. By removing the influence of the measuring chain (measuring instruments, especially detectors), corrected amplitudes and phases (black line) are obtained which can be analyzed by equations 1-4 by the standard fitting method. The results of such analysis of the corrected signal give values of silicon $(l_2 = 30\ \mu m)$ which corresponds to standard silicon substrate (thin plate) thicknesses, titanium-dioxide $(l_1 = 500\ nm)$ which corresponds to standard thin-film thicknesses, radius $R = 3$ mm, while other parameters correspond to the parameters from Table 1 with an error of 5%. The corrected signals from Figure 5 are further presented to our networks and the results of their prediction are given in Table 6 and Table 7. The relative error in these tables is the result of comparing network predictions and standard fitting of the existing theoretical model.

**Table 6.** Parameters $l_1^{NN1}$, $\alpha_{T1}^{NN2}$ and $D_{T1}^{NN3}$ obtained by prediction of NN1-3, with relative (%) errors are calculated according to the parameters obtained from standard photoacoustics techniques.

| parameter | $l_1^{NN1}$ | $\alpha_{T1}^{NN2}$ | $D_{T1}^{NN3}$ |
|---|---|---|---|
| NN exp prediction | 4.8018x10$^2$ nm | 1.0955x10$^{-5}$ K$^{-1}$ | 3.57913x10$^{-6}$ m$^2$s$^{-1}$ |
| relative (%) error | 3.9644 | 0.4066 | 2.9372 |

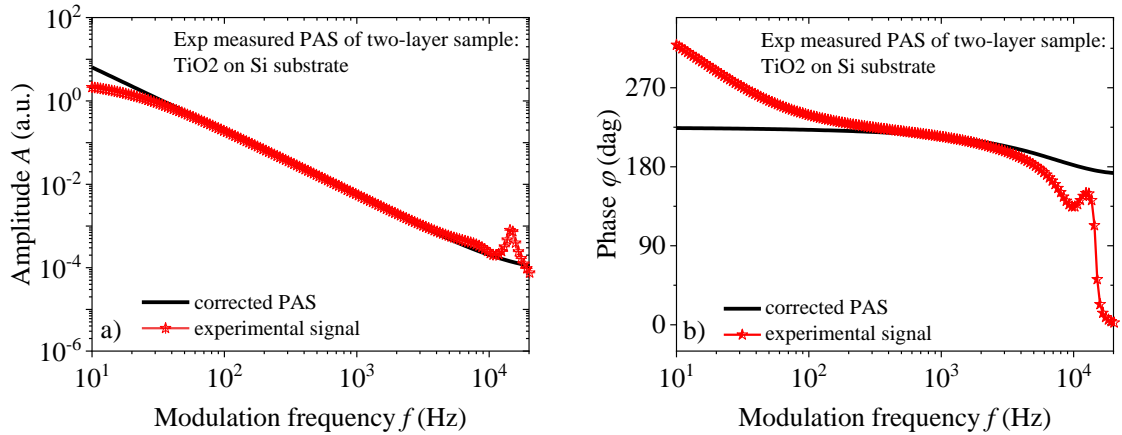

Figure 5. Frequency dependence of a) amplitude and b) phase of experimentally measured PAS TiO$_2$ placed on Si substrate (red asterisk) and the corresponding amplitude and phase of the PAS $\delta p_{total}(f)$ corrected signal on the instrument input (black line).

Based on the results of prediction by neural networks NN1-3, (Table 6), the most accurate network is NN2 in the prediction of the thermal expansion coefficient $\alpha_{T1}^{NN2}$ of a thin-film TiO$_2$, with a relative (%) error <1%, while the precision in the prediction of the thermal diffusivity $D_{T1}^{NN3}$ and thickness $l_1^{NN1}$ is with relative (%) errors <5%.

**Table 7.** Parameters $l_1^{NN4}$, $\alpha_{T1}^{NN4}$ and $D_{T1}^{NN4}$ obtained by prediction of NN4, with relative (%) errors are calculated according to the parameters obtained from standard photoacoustics techniques.

| parameter | $l_1^{NN4}$ | $\alpha_{T1}^{NN4}$ | $D_{T1}^{NN4}$ |
|---|---|---|---|
| NN4 exp prediction | 4.8690x10$^2$ nm | 1.1166x10$^{-5}$ K$^{-1}$ | 3.7189x10$^{-6}$ m$^2$s$^{-1}$ |
| relative (%) error | 2.6196 | 1.5106 | 0.5105 |

In the simultaneous prediction of the parameters of thickness $l_1^{NN4}$, thermal expansion coefficient $\alpha_{T1}^{NN4}$ and thermal diffusivity $D_{T1}^{NN4}$ (Table 7), the NN4 network gives satisfactory results comparable to the prediction results of NN1-3.

Despite the expectations based on the consideration of the theoretical model, which is reflected in the small visual difference of the amplitude characteristics and stratification of signal phases in the high-frequency part (1-20) kHz, neural networks based on the coupled amplitudes and phases in the frequency domain (20-20k) Hz, can determine the parameters of the thin-layer TiO$_2$. The results of neural networks show that more precise and accurate results are obtained in networks in which multiple parameters are determined at the same time (Tables 3 and 5) than in networks in the prediction of individual parameters (Tables 2 and 4). This conclusion is also valid for the prediction of the thin-film parameter from the experimental results, where the reduction of the relative % error in the prediction of the network NN4 in relation to NN1-3 is observed, which can represent one of the methods of optimizing the work of networks in prediction the parameter of thin-films.

This consideration is particularly valid due to the analysis of a thin-layer of TiO$_2$ placed on a well-characterized substrate, in this case silicon. The method of characterization of TiO$_2$ developed in this way can be applied and analyzed on other well-characterized optically transparent and non-transparent substrates. By applying TiO$_2$ to optically transparent substrates, and by characterizing it, we obtain a suitable material for protecting the detectors of the measuring system.

**Conclusions**

The results presented in this paper indicate one very important fact - neural networks are able to easily recognize changes in thin-films deposited on a substrate even though the thicknesses of such films are in the submicron domain (limit of detection for most PA methods).

Theoretical analyzes of two-layer samples Si (substrate) + $TiO_2$ (thin-film) showed relatively easy recognition of changes in film thickness of ±5 nm, coefficient of thermal expansion of ±5 $10^{-8}$ $K^{-1}$ and coefficient of thermal diffusion of ±18.5 $10^{-8}$ $m^2s^{-1}$.

In addition, it has been shown that neural networks for predicting thin-film parameters can be well trained with a relatively small database. It has been shown that even on such small bases, networks can be formed that will give a prediction of only one parameter, but also networks that give a prediction of three parameters simultaneously. It has been shown that all networks give approximately the same level of prediction both in theoretical considerations and in the analysis of experimental data. Therefore, it can be recommended that for the analysis of thin-films on different substrates, it is enough to form one network that simultaneously predicts several of its parameters. It has been shown that the formation of several networks that predict only one parameter is unpractical.

**Acknowledgments**

We are thankful for the financial support of this research by the Ministry of Education, Science and Technology development of the Republic of Serbia, contract number 451-03-09/2021-14/200017.

**Authors declarations** Conflict of Interest The authors have no conflicts to disclose.

**Data Availability** The data that support the findings of this study are available from the corresponding author upon reasonable request.

**Appendix I. Temperature distributions in two-layer sample**

Periodic temperature distributions in the thin-film (1-TiO$_2$) and substrate (2-Si) illuminated by the modulated light source (Figure 1) can be obtained solving the thermal-diffusion equations in the form [23,52,55]:

$$\frac{\partial^2 T_1(z,f)}{\partial z^2} - \frac{i\omega}{D_{T1}} T_1(z,f) = -\frac{1}{k_1} \beta_1 (1-R_1) I_0 e^{-\beta_1 z}, \qquad (20)$$

and

$$\frac{\partial^2 T_2(z,f)}{\partial z^2} - \sigma_2^2 T_2(z,f) = -\frac{\varepsilon_g}{k_2 \tau_2} n_{p2}(z,f) - \frac{\beta_2 I}{k_2} \cdot \frac{\varepsilon - \varepsilon_g}{\varepsilon} e^{-\beta_2 z}, \qquad (21)$$

where $\omega = 2\pi f$, $f$ is the modulation frequency, $I_0$ is the incident light intensity, $I = (1-R_1)(1-R_2)e^{-\beta_1 l_1} I_0$, $R_1$ is the film reflection coefficient, $R_2$ is the substrate reflection coefficient, $\sigma_1 = \sqrt{i\omega/D_{T1}}$ is the film complex thermal diffusivity, $D_{T1}$ is the film thermal diffusion coefficient, $\sigma_2 = \sqrt{i\omega/D_{T2}}$ is the substrate complex thermal diffusivity, $D_{T2}$ is the substrate thermal diffusion coefficient, $k_1$ is the thin-film heat conduction coefficient, $k_2$ is the substrate heat conduction coefficient, $\beta_1$ is the film absorption coefficient, $\beta_2$ is the substrate absorption coefficient, and $\delta n_{p2}(z,f)$ is the substrate photogenerated minority carrier dynamic density component (Eq.(18)).

The general solutions of Eqs.(17,18) can be written in the form [23,52,55]:

$$T_1(z,f) = A_1 e^{\sigma_1 z} + A_2 e^{-\sigma_1 z} + A_3 e^{-\beta_1 z}, \qquad (22)$$

and

$$T_2(z,f) = B_1 e^{\sigma_2 z} + B_2 e^{-\sigma_2 z} + B_3 \delta n_p(z,f) + B_4 e^{-\beta_1 z}, \qquad (23)$$

where the constants $A_3$, $B_3$ and $B_4$ are given as:

$$A_3 = -\frac{\beta_1 I_0 (1-R_1)}{k_1 (\beta_1^2 - \sigma_1^2)}, \quad B_3 = -\frac{\varepsilon_g}{k_2 \tau_{p2} \left(\sigma_2^2 - \dfrac{1}{L_2^2}\right)},$$

$$B_4 = -\frac{\beta_2 (1-R_1)(1-R_2) e^{-\beta_1 l_1} I_0}{\varepsilon(\beta_2^2 - \sigma_2^2)} \left(\frac{B_3}{D_{p2}} - \frac{\varepsilon - \varepsilon_g}{k_2}\right).$$

Here $L_2 = \sqrt{\dfrac{D_{p2} \tau_{p2}}{1 + i\omega \tau_{p2}}}$ is the complex minority carrier diffusion length, $D_{p2}$ is the diffusion coefficient of minority carriers (holes $p$ in the $n$-type substrate), and $\tau_{p2}$ is the bulk minority carrier lifetime. Constants $A_1$, $A_2$, $B_1$ and $B_2$ can be found solving the boundary conditions [23,52,55]:

$$\text{a) } -k_1 \left.\frac{\partial T_1(z,f)}{\partial z}\right|_{z=-l_1} = 0, \quad \text{b) } T_1(0,f) = T_2(0,f),$$

$$\text{c) } -k_2 \left.\frac{\partial T_2(z,f)}{\partial z}\right|_{z=0} = s_F n_{p2}(0,f)\varepsilon_g - k_1 \left.\frac{\partial T_1(z,f)}{\partial z}\right|_{z=0},$$

$$\text{d) } -k_2 \left.\frac{\partial T_2(z,f)}{\partial z}\right|_{z=l_2} = -s_R n_{p2}(l_2,f)\varepsilon_g, \tag{24}$$

where $s_F$ and $s_R$ are the substrate surface recombination speeds at the front $(z=0)$ and rear $(z=l_2)$ surfaces, respectively.

Based on our previous investigations [23,49,52,55], the analysis of the two-layer optical properties shows that the multiple optical reflections can be neglected in the Si substrate [23], but must be taken into account in the case of thin TiO$_2$ film. This is the reason why the film reflection coefficient $R_1$ is calculated here using [52,55]:

$$R_1 = r_F + (1-r_F)^2 r_R \cdot \frac{e^{-2\beta_1 l_1}}{1 - r_F r_R e^{-2\beta_1 l_1}}, \tag{25}$$

where $r_F$ and $r_R$ are the front and rear thin-film reflectivity coefficients, respectively.

**Appendix II. Two-layer sample displacement along the heat-flow axes**

The $U_{z,c}(r,z)$ of the two-layer sample at the back surface, $z = l_2$, important in a transmission photoacoustic measurements, can be written in a general form as [23,52,55]:

$$U_{z,c}(r,z) = \frac{C_c}{2}\left(R_s^2 - r^2\right), \quad c = \text{TE, PE}, \tag{4}$$

where $R_s$ is the sample radius and

$$C_{TE} = 6 \frac{A_1 + A_2 + E_1 E_2 \left[ \alpha_{T1} l_2 (2M_{T1} - l_2 N_{T1}) + \alpha_{T2} l_1 (2M_{T2} + l_1 N_{T2}) \right]}{E_1^2 l_1^4 + E_2^2 l_2^4 + 2 E_2 E_1 l_2 l_1 (2 l_2^2 + 3 l_2 l_1 + 2 l_1^2)}, \qquad (5.a)$$

$$C_{PE} = 6 d_n E_2 \frac{\left[ E_1 l_1 (2M_n + l_1 N_n) + E_2 l_2 (2M_n - l_2 N_n) \right]}{E_2^2 l_2^4 + E_1^2 l_1^4 + 2 E_2 E_1 l_2 l_1 (2 l_2^2 + 3 l_2 l_1 + 2 l_1^2)}. \qquad (5.b)$$

Here $A_1 = E_1^2 l_1 (2M_{T1} + l_1 N_{T1}) \alpha_{T1}$, $A_2 = E_2^2 l_2 (2M_{T2} - l_2 N_{T2}) \alpha_{T2}$, $E_1$ and $E_2$ are the Young's modulus of the film and substrate, respectively, $d_n$ is the coefficient of electronic deformation and $M_{T1}$, $M_{T2}$, $M_n$, $N_{T1}$, $N_{T2}$ and $N_n$ are defined as [23,52,55]

$$M_{T1} = \int_{-l_1}^{0} z \cdot T_1(z, f) dz, \quad M_{T2} = \int_{0}^{l_2} z \cdot T_2(z, f) dz, \quad M_n = \int_{0}^{l_2} z \cdot \delta n_{p2}(z, f) dz, \qquad (6)$$

$$N_{T1} = \int_{-l_1}^{0} T_1(z, f) dz. \quad N_{T2} = \int_{0}^{l_2} T_2(z, f) dz, \quad N_n = \int_{0}^{l_2} \delta n_{p2}(z, f) dz, \qquad (7)$$

where $T_1(z, f)$ is the temperature in the thin-film and $T_2(z, f)$ is the temperature in the substrate and $\delta n_{p2}(z, f)$ is the photogenerated minority carrier density. The $M_{T1}$, and $M_{T2}$ are the first moments of the temperature change, and the $M_n$ is the first moment of the photogenerated minority carriers change along the z-axis. The $N_{T1}$ and $N_{T2}$ are the average temperature changes and $N_n$ is the average photogenerated minority carriers change along the z-axes.